\documentclass[twocolumn,prl,showpacs,preprintnumbers,floatfix]{revtex4}%
\usepackage{amssymb}
\usepackage{graphicx}
\usepackage{dcolumn}
\usepackage{bm}
\usepackage{longtable}
\usepackage{amsmath}
\usepackage{amsfonts}
\usepackage{booktabs}%
\begin{document}
\title{A new scissors mode on the skin of deformed neutron rich nuclei}
\author{D. Pena Arteaga$^{1}$}%
\author{P. Ring$^{2,3}$}
\affiliation{$^1$ Institut de Physique Nucl\'eaire, Universit\'e
Paris-Sud, IN2P3-CNRS, F-91406 Orsay Cedex, France}%
\affiliation{$^2$ Physikdepartment, Technische Universit\"{a}t M\"{u}nchen,
D-85748, Garching, Germany}%
\affiliation{$^3$ Departamento de F\'{\i}sica Te\'{o}rica, Universidad
Aut\'{o}noma de Madrid, E-28049 Madrid, Spain}
%\date{\today}

\begin{abstract}
Covariant density functional theory is used to analyze the evolution
of low-lying M1 strength in superfluid deformed nuclei in the
framework of the self-consistent Relativistic Quasiparticle Random
Phase Approximation (RQRPA). In nuclei with a pronounced neutron
excess two scissor modes are found. Besides the conventional scissor
mode, where the deformed proton and neutron distributions oscillate
against each other, a new soft M1 mode is found, where the deformed
neutron skin oscillates in a scissor like motion against a deformed
proton-neutron core.

\end{abstract}

\maketitle

Since the discovery of the of the Giant Dipole Resonance (GDR) in the
photo absorption spectra of nuclei and its interpretation as an
oscillation of protons against neutrons~\cite{GT.48}, many other
collective excitations have been found, and much has been learned
about the bulk properties of nuclei from such resonances. The
interpretation of such modes is closely linked with the symmetries of
the underlying system, and the phase transitions connected with their
spontaneous breaking. The simplest example is the translational
motion of the entire nucleus, protons and neutrons in phase, with the
quantum numbers $J^{\pi}=1^{-},$ $T=0$. Because of translational
invariance, this mode is not really an excitation, but it shows up in
the theoretical excitation spectra as a Goldstone mode connected with
the symmetry violation by the mean field approach. Closely related is
the vibration of neutrons against protons, with the quantum numbers
$J^{\pi}=1^{-},$ $T=1$. It is, of course, not spurious and
corresponds to the Giant Dipole Resonance (GDR).

A similar pattern has been observed in deformed nuclei. Rotational
symmetry is spontaneously broken and the corresponding Goldstone mode
with the quantum numbers $K^{\pi}=1^{+},$ $T=0$ in the intrinsic
system is connected with the collective rotation of protons and
neutrons in-phase. Since thirty years it is also known that there is
a vibrational mode with the quantum numbers $K^{\pi }=1^{+},$ $T=1$,
corresponding to a scissor-like rotational motion of neutrons against
protons~\cite{SR.77a,IP.78,Iac.81,BRS.84x,Hil.84}. It can be excited
by the magnetic dipole (M1) operator and has been found in many
nuclei at energies of a few MeV by scattering of proton, electrons or
photons (for reviews and recent applications see
Refs.~\cite{Rich.95,KPZ.96x,BalS.07a}).

In recent years, new experimental facilities have permitted the study
a variety of new phenomena connected with the isospin degree of
freedom. Systems with large neutron excess show a pronounced neutron
skin and much experimental and theoretical interest has been devoted
to the study of collective modes connected with this skin. The Pygmy
Dipole Resonance (PDR) identified in the E1-strength distribution at
low energies has been interpreted as a resonant oscillation of the
neutron skin against the remaining isospin saturated neutron-proton
core. It has been observed not only in light systems \cite{Lei.01},
but also in heavy nuclei with neutron excess in and far from the
valley of $\beta$-stability~\cite{Adr.05}. This mode is also expected
to play an important role in astrophysical applications in neutron
rich systems, where the presence of a low-lying resonant component of
the E1 strength has a strong influence on the radiative
neutron-capture rates in the r-process~\cite{Gor.98}.

On the theoretical side, the pygmy mode was first predicted in
hydrodynamical models~\cite{MDB.71x,SIS.90x} and, later, in the
context of density functional theory~\cite{Rei.99}. More recently, it
has been studied with the aid of RPA and QRPA calculations, as well
as large scale shell model applications (for a recent review see
Ref.~\cite{PVK.07x} and references given therein). Since the pygmy
mode has translational character, its quantum numbers are
$J^{\pi}=1^{-}$. The isospin is mixed. In the nuclear interior
protons and neutrons move in phase, forming a $T=0$ core, while near
the nuclear surface, build up predominantly by neutrons, one has a
superposition of $T=0$ and $T=1$.

In the PDR, the neutron skin describes a translational motion with
respect to a core composed of neutrons and protons. It is natural to
expect that, in a similar way, the skin plays also a role in the
scissors mode in deformed nuclei with rotational character. This has
been investigated in Ref.~\cite{WI.97x} in a boson model with three
degrees of freedom (protons, core neutrons and neutron skin) and it
has been shown that group theory allows several possible modes. One
of them is the normal scissors mode, where the protons oscillate
against the combined system of neutrons (core and skin). In addition
a new mode is possible, where the protons form together with the
neutrons a deformed core system with $T=0$, which oscillates in a
scissor like motion against the deformed neutron skin. However, there
are further modes possible and in a group theoretical model it is
hard to decide which of them is realized in actual nuclei.

In the present manuscript we investigate this question in a fully
microscopic and self-consistent way using covariant density
functional theory. As an example we concentrate on the nucleus
$^{154}$Sm, which has an excess of 30 neutrons. Density functional
theory based on the mean field approach plays an important role in a
fully microscopic and universal description of nuclei all over the
periodic table. Covariant density functional theory (CDFT) is
particularly successful because Lorentz invariance reduces the number
of parameters considerably. Pairing correlations can be included in
the framework of relativistic Hartree-Bogoliubov (RHB) theory. For
recent reviews see Refs.~\cite{Rin.96,VALR.05}. Starting from the
time-dependent version of CDFT, the same functionals can also be
applied to investigate nuclear excitations. In the small amplitude
limit one finds the relativistic Random Phase Approximation
(RRPA)~\cite{RMG.01x}, suited for the description of excited states
with vibrational character. It has been used with great success for
the study of giant resonances and spin- or/and isospin-excitations as
the Gamow-Teller Resonance (GTR) or the Isobaric Analog Resonance
(IAR)~\cite{PNV.04}. Recently it also has been applied for a
theoretical interpretation of the low lying E1
strength~\cite{PRN.03}.

So far, applications or relativistic RPA and QRPA have been
restricted to spherical systems. Here we report on the first
theoretical investigation using a fully self-consistent
implementation of relativistic QRPA for deformed nuclei with axial
symmetry. We use the parameter set NL3~\cite{LKR.97} , which has been
thoroughly tested and proven to successfully describe many nuclear
properties. Pairing correlations are taken into account using a
monopole pairing force with the strength parameters adjusted to the
experimental pairing gaps obtained from odd-even mass differences.

First the relativistic mean field equations are solved in the basis
of an anisotropic axial harmonic oscillator with 20 major
shells~\cite{GRT.90x}. The resulting Dirac spinors for the
ground-state are subsequently used to construct two-quasiparticle
pairs with the appropriate quantum numbers $K^{\pi}=1^{+}$, and to
evaluate the matrix elements of the effective interaction obtained
from the second derivative of the original energy functional using a
Fourier-Bessel decomposition in cylindrical
coordinates~\cite{Pen.07,PR.08}. No new parameters are necessary. The
solution of the QRPA equations provide the excitation spectrum, and
allows the calculation of the dynamical M1-response $R_{M1}(\omega)$
and the corresponding strength function. Since the starting point is
a deformed mean-field the calculated response function refers to the
intrinsic frame. It has to be transformed to the laboratory frame by
projection onto good angular momentum $I=1$ using the needle
approximation~\cite{RRE.84x}, which is valid for well-deformed
nuclei.

The spontaneous breaking of rotational symmetry in the ground-state
leads to a spurious $K^{\pi}=1^{+}$ excitation at zero energy in the
QRPA spectrum which corresponds to a rotation perpendicular to the
symmetry axis. One the advantages of self-consistent QRPA is that, to
the extent numerical errors are kept to a minimum, this spurious mode
decouples exactly at zero energy. By taking into account all the
mesons and their currents, as well as the electromagnetic fields, in
the QRPA matrix elements, and by using a large two-quasiparticle
space, the position of the spurious mode is in our calculations below
$0.5$ MeV, and this excitation exhausts more than 99\% of the total
spurious $J_{\pm1}$ strength in $^{154}$Sm, i.e. its admixture with
physical states is negligible.

\begin{figure}[ptbh]
\includegraphics[width=8.0cm]{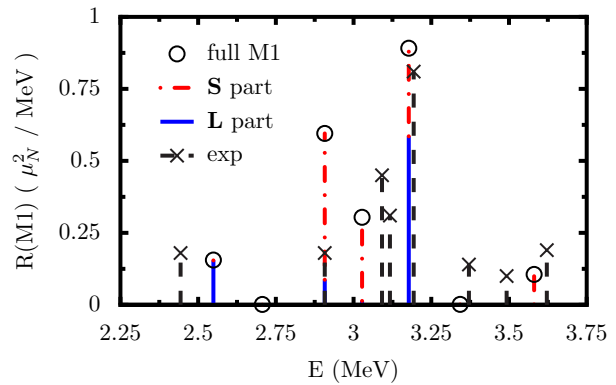}\caption{(Color online) M1 transition
strength in $^{154}$Sm. Dashed lines with crosses show the experimentally
known M1 excitations \cite{ZHN.93x}. The full and dot-dashed lines (black)
with circles mark the theoretical RQRPA response. In these peaks, the
contribution coming from the angular part of the M1 operator is represented by
the full line (blue), while the dot-dashed lines (red) belong to the spin
response.}%
\label{fig1}\end{figure}

In Fig.~\ref{fig1} we show the low-lying M1 response in $^{154}$Sm
obtained within RQRPA together with the experimentally known
excitations \cite{ZHN.93x}. The agreement is remarkable. In
particular, the theoretically determined position and strength of the
strongest peak at 3.2~MeV, and its orbital nature, coincide with the
established experimental knowledge. Four out of the six most
prominent theoretical peaks have mostly spin character. We
concentrate in the following on the two remaining modes with orbital
character. Besides the scissor mode at 3.2~MeV a second low-lying
orbital excitation is found around 2.5~MeV. These two excitations
with orbital nature can be interpreted in geometrical terms by
analyzing their intrinsic transition densities $\delta\rho_{\nu}$ in
cylindrical coordinates $r,z$ and $\varphi$, $r$ being the distance
from the symmetry axis in $z$-direction. $\delta\rho_{\nu}$ is
related, in the harmonic approximation, to the time-dependent density
distribution for a given excitation mode $\nu$ with quantum numbers
$K^{\pi}$

\begin{figure}[ptbh]
\includegraphics[width=8.0cm]{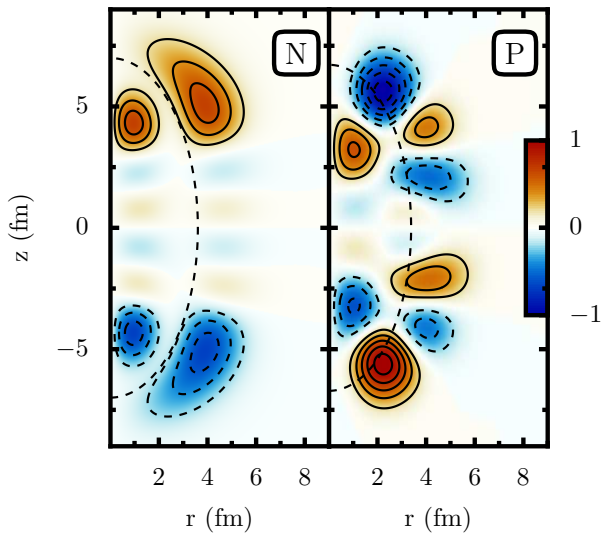}\caption{(Color online) Intrinsic
transition density for neutrons (left) and protons (right) of the M1
peak at 3.2~MeV in $^{154}$Sm. Full and dashed lines indicate
positive and negative values, respectively. The $z$ coordinate runs
along the symmetry axis. The thin dotted line represents the rms
radius of the ground state neutron or proton density.} \label{fig2}
\end{figure}

\begin{equation}
\rho(\mathbf{r},t)=\rho^{(0)}(r,z)+
{\displaystyle\sum\limits_{\nu}}
(\delta\rho_{\nu}(r,z)e^{iK\varphi}e^{-i\omega_{\nu}t}+\mathrm{h.c.),}%
\end{equation}
where $\rho^{(0)}$ is the axially symmetric ground state density and
$\hslash\omega_{\nu}$ are the peak energies of the various excitations
$|\nu\rangle$. The quantities $\delta\rho_{\nu}$ are the so-called intrinsic
transition densities. These quantities provide an intuitive understanding of
the geometrical nature of the excitation modes and they shall be used in the
following for a discussion of their internal structure.

\begin{figure}[ptbh]
\includegraphics[width=8.0cm]{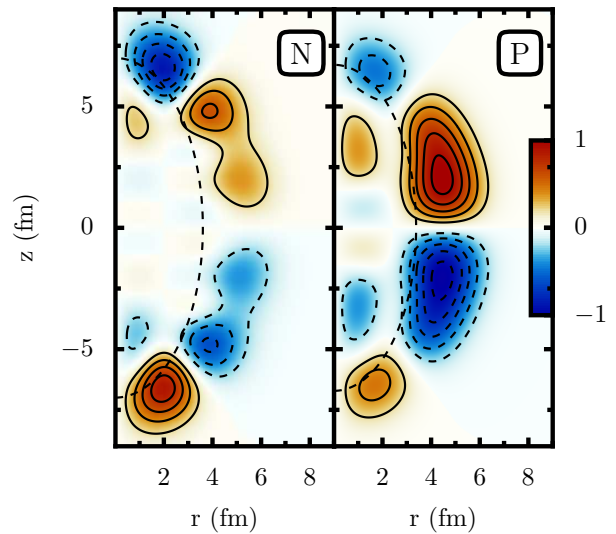}\caption{(Color online) Intrinsic
transition density of the M1 peak at 2.5 MeV in $^{154}$Sm (see Fig.
2 and
text for details).}%
\label{fig3}%
\end{figure}

\begin{figure}[h]
\includegraphics[width=8.0cm]{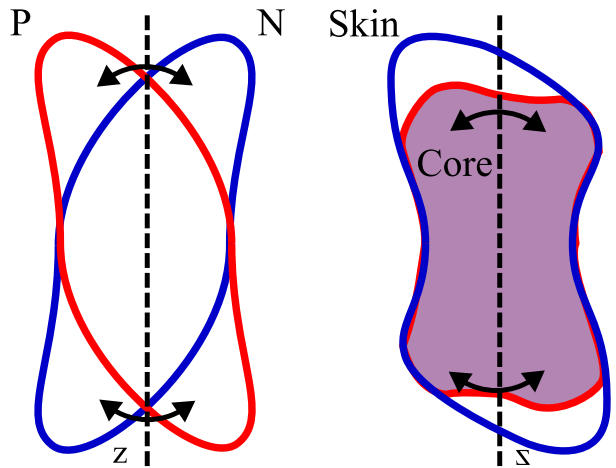}\caption{(Color online) Schematic
drawing of the two low-lying orbital modes. On the left is the
typically scissors-like motion, which corresponds to the peak at
3.86~MeV. The
right-hand side corresponds to the low lying peak at 2.60~MeV. }%
\label{fig4}%
\end{figure}

Figures ~\ref{fig2} and~\ref{fig3} show the intrinsic transition
densities $\delta\rho_{\nu}(r,z)$ for the two orbital peaks at
$\hslash\omega_{\nu}=3.2$~~MeV and at $\hslash\omega_{\nu}=2.5$ MeV
as contour plots in the ($r,z$)-plane. In Fig.~\ref{fig2} we observe
the typical structure of the normal scissors mode. Neutrons and
protons are out of phase over most of the space, with a concentration
near the caps ($z>5$ fm) of the prolate nuclear shape. Effects of the
mixing of orbital and spin strength can be seen in the region around
$z\approx2.5$ fm and $r\approx4$ fm in the proton intrinsic
transition density. A rough schematic representation of such a mode
is shown on the left hand side of Fig.~\ref{fig4}.

On the other hand, the low lying mode at 2.5~MeV, whose intrinsic
transition density is plotted in Fig.~\ref{fig3}, shows a completely
different spatial excitation pattern. Like the scissors mode, it is
mostly a surface vibration. However, unlike the normal scissors mode
in Fig.~\ref{fig2}, which is clearly a $T=1$ mode, its isospin is
mixed, closer to being $T=0$: over the entire area protons and
neutrons are in phase. However there is a clear distinction between
the outer part of the nucleus with $z>5$ fm representing the skin
 and the inner region with $z<5$ fm representing the core. These two parts
clearly oscillate against each other. We also find that the skin consists
mainly of neutrons. In a first approximation we therefore have an neutron skin
oscillating against the proton-neutron core. Of course the details are more
complicated and there are small admixtures of protons in the skin.

The schematic representation on the right side of Fig.~\ref{fig4}
portrays such a situation. The shaded region ($z<5$~fm) represents
the nuclear core composed of neutrons and protons rotating in phase.
The full line represents the extended mostly neutron skin at the cap
which rotates out of phase against the core.

The quasiparticle structure found in QRPA provides detailed
information on the character of the corresponding mode. In
Table~\ref{table1} we show the decomposition of the two low-lying
orbital modes discussed above into their quasiparticle components.
Each of the two Dirac spinors in the quasiparticle pair corresponds
to a eigen state of the static RMF potential and can be characterized
by the Nilsson quantum numbers $\Omega^{\pi}[Nn_{z}\Lambda]$ of its
largest component in an expansion in anisotropic oscillator wave
functions. Here $\Omega$ is the total angular momentum projection
onto the symmetry $z$-axis, $\pi$ is the parity,
$N=2n_{r}+n_{z}+\Lambda$ is the major oscillator quantum number, and
$\Lambda=\Omega-m_{s}$ is the projection of the orbital angular
momentum onto the symmetry axis.

\begin{table}[ptb]
\begin{center}%
\begin{tabular}
[c]{lrlcrc}
&  & \multicolumn{3}{c}{Peak at 2.5~MeV} & $E_{1}+E_{2}$\\
P & 58\% & $\frac{5}{2}^{+}[413]$ & $-$ & $\frac{7}{2}^{+}[404]$ & 3.13\\
N & 20\% & $\frac{3}{2}^{+}[651]$ & $-$ & $\frac{5}{2}^{+}[642]$ & 2.75\\
N & 3\% & $\frac{3}{2}^{-}[532]$ & $-$ & $\frac{5}{2}^{-}[523]$ & 3.64\\
P & 3\% & $\frac{3}{2}^{-}[541]$ & $-$ & $\frac{5}{2}^{-}[532]$ & 3.10\\
N & 3\% & $\frac{1}{2}^{+}[660]$ & $-$ & $\frac{3}{2}^{+}[651]$ & 3.05\\
\bottomrule &  &  &  &  & \\
&  & \multicolumn{3}{c}{Peak at 3.2~MeV} & $E_{1}+E_{2}$\\
P & 65\% & $\frac{3}{2}^{-}[541]$ & $-$ & $\frac{5}{2}^{-}[532]$ & 3.10\\
N & 9\% & $\frac{3}{2}^{-}[532]$ & $-$ & $\frac{5}{2}^{-}[523]$ & 3.64\\
P & 7\% & $\frac{5}{2}^{+}[413]$ & $-$ & $\frac{7}{2}^{+}[404]$ & 3.13\\
N & 7\% & $\frac{1}{2}^{-}[530]$ & $-$ & $\frac{3}{2}^{-}[521]$ & 3.66\\
P & 4\% & $\frac{1}{2}^{+}[420]$ & $-$ & $\frac{3}{2}^{+}[411]$ & 4.37\\
&  &  &  &  &
\end{tabular}
\end{center}
\caption{Quasiparticle structure of the two lower QRPA excitation modes with
orbital character in the M1 response for $^{154}$Sm. N and P indicate a
neutron or proton quasiparticle pair, respectively. The second column is the
percentage of the contribution of each particular two-$qp$ excitation. The
Nilsson quantum numbers labeling the quasiparticle are shown in the next
columns. The last column gives the energy of the $qp$-excitation in MeV.}%
\label{table1}%
\end{table}

From these quantum numbers one concludes the following approximate
selection rules for both orbital modes: $\Delta\Omega=+1$, $\Delta
N=0$, $\Delta n_{z}=-1$ and $\Delta\Lambda=+1$. Their orbital
character is confirmed by the fact that $\Delta\Omega=\Delta\Lambda$,
which implies that the change in the spin quantum number $m_{s}$ is
zero. Both modes are distributed upon the same set of $qp$
excitations, differing mainly in their relative admixture. While the
peak at 3.2~MeV is evidently more equally distributed amongst protons
and neutrons indicating an isovector excitation with $T=1$, the lower
lying excitation at 2.5~MeV is a combination of both isoscalar $T=0$
and $T=1$ isovector.

In summary, a new code has been developed to solve the relativistic
QRPA equations for axially deformed systems \cite{Pen.07,PR.08}.
Using covariant density functional theory based on the parameter set
NL3~\cite{LKR.97} the low-lying orbital magnetic response has been
investigated for the nucleus $^{154}$Sm. In agreement with
experimental data \cite{ZHN.93x}, it has been found that in this
nucleus the well known scissors mode is split into several
resonances, two of them having mostly orbital character. The
calculations show that the upper peak at 3.2 MeV corresponds to the
normal scissors mode with protons oscillating against neutrons,
whereas the lower peak at 2.5 MeV corresponds to a new type of
resonance: the loosely bound neutron skin oscillates in a scissor
like motion against the proton-neutron core with pronounced $T=0$
character. This provides a new understanding of the splitting of low
lying orbital $K=1^{+}$ modes observed in many deformed nuclei.
Further investigations of this mode in nuclei of the periodic table
are in progress. In particular it has to be clarify to which extend
this mode depends on the underlying single particle structure.

\begin{acknowledgments}
Helpful discussions with D. Vretenar are gratefully acknowledged. P.R. thanks
for the support provided by the Ministerio de Educaci\'on y Ciencia, Spain.
The paper has also been supported
%by the Bundesministerium f\"{u}r Bildung und
%Forschung, Germany under project 06 MT 246 and
by the DFG cluster of excellence \textquotedblleft Origin and
Structure of the Universe\textquotedblright\
(www.universe-cluster.de).
\end{acknowledgments}

\vspace{0.3cm}

%\bibliographystyle{c:/E/a00/prsty}
%\bibliography{c:/B/a00/refring}

\begin{thebibliography}{10}

\bibitem{GT.48}
M. Goldhaber and E. Teller, Phys. Rev. {\bf 74},  1046  (1948).

\bibitem{SR.77a}
T. Suzuki and D.~J. Rowe, Nucl. Phys. {\bf A289},  461  (1977).

\bibitem{IP.78}
N. LoIudice and F. Palumbo, Phys. Rev. Lett. {\bf 41},  1532
(1978).

\bibitem{Iac.81}
F. Iachello, Nucl. Phys. {\bf A358},  89c  (1981).

\bibitem{Hil.84}
R.~R. Hilton, Z. Phys. {\bf A316},  121  (1984).

\bibitem{BRS.84x}
D. Bohle~{\it et al}, Phys. Lett. {\bf B137},  27  (1984).

\bibitem{Rich.95}
A. Richter, Progr. Part. Nucl. Phys. {\bf 34},  261  (1995).

\bibitem{KPZ.96x}
U. Kneissl~{\it et al.}, Progr. Part. Nucl. Phys. {\bf 37},  349
(1996).

\bibitem{BalS.07a}
E.~B. Balbutsev and P. Schuck, Ann. Phys. (N.Y.) {\bf 322},  489
(2007).

\bibitem{Lei.01}
A. Leistenschneider~{\it et al.}, Phys. Rev. Lett. {\bf 86},  5442
(2001).

\bibitem{Adr.05}
P. Adrich~{\it et al.}, Phys. Rev. Lett. {\bf 95},  132501  (2005).

\bibitem{Gor.98}
S. Goriely, Phys. Lett. {\bf B436},  10  (1998).

\bibitem{MDB.71x}
R. Mohan~{\it et al.}, Phys. Rev. {\bf C3},  1740  (1971).

\bibitem{SIS.90x}
Y. Suzuki~{\it et al.}, Prog. Theor. Phys. {\bf 83},  180  (1990).

\bibitem{Rei.99}
P.-G. Reinhard, Nucl. Phys. {\bf A649},  305c  (1999).

\bibitem{PVK.07x}
N. Paar~{\it et al.}, Rep. Prog. Phys. {\bf 70},  691  (2007).

\bibitem{WI.97x}
D.~D. Warner~{\it et al.}, Phys. Lett. {\bf B395},  145  (1997).

\bibitem{Rin.96}
P. Ring, Prog. Part. Nucl. Phys. {\bf 37},  193  (1996).

\bibitem{VALR.05}
D. Vretenar, A.~V. Afanasjev, G.~A. Lalazissis, and P. Ring, Phys.
Rep. {\bf
  409},  101  (2005).

\bibitem{RMG.01x}
P. Ring~{\it et al.}, Nucl. Phys. {\bf A694},  249  (2001).

\bibitem{PNV.04}
N. Paar, T. Nik\v{s}i\'{c}, D. Vretenar, and P. Ring, Phys. Rev.
{\bf C69},
  054303  (2004).

\bibitem{PRN.03}
N. Paar, P. Ring, T. Nik{\v{s}}i{\'{c}}, and D. Vretenar, Phys. Rev.
{\bf C67},
   034312  (2003).

\bibitem{LKR.97}
G.~A. Lalazissis, J. K\"{o}nig, and P. Ring, Phys. Rev. {\bf C55},
540
  (1997).

\bibitem{GRT.90x}
Y.~K. Gambhir~{\it et al.}, Ann. Phys. (N.Y.) {\bf 198},  132
(1990).

\bibitem{Pen.07}
D.~P. Arteaga, Phd thesis, Technische Universit\"{a}t M\"{u}nchen
  (unpublished), 2007.

\bibitem{PR.08}
D.~P. Arteaga and P. Ring, to be published.

\bibitem{RRE.84x}
P. Ring~{\it et al.}, Nucl. Phys. {\bf A419},  261  (1984).

\bibitem{ZHN.93x}
W. Ziegler~{\it et al.}, Nucl. Phys. {\bf A564},  366  (1993).

\end{thebibliography}

\end{document}